\documentclass{article}[11pt]
\usepackage{epsfig}
\usepackage{amsmath}
\usepackage{amsfonts}
\usepackage{color}

\def\baselinestretch{1}  

\begin{document}
\title{Atomic clock with a nuclear transition:\\ current status in TU Wien}
    
\author{G. Kazakov$^{1,2}$, M. Schreitl$^1$, G. Winkler$^1$, \\
J. H. Sterba$^1$, G. Steinhauser$^1$, and T. Schumm$^1$ \\
\\
\multicolumn{1}{p{.85\textwidth}}{\centering\emph{
$^{1}$ Institute of Atomic and Subatomic Physics TU Wien, Stadionallee 2, Vienna 1020, Austria \\
\vspace{1mm}
$^{2}$ St. Petersburg State Polytechnical University, 29, Polytechnicheskaya st, 
St. Petersburg 195251, Russia}}}
\date{}
\maketitle

\begin{abstract}
The nucleus of $^{229}$Thorium presents a unique isomer state of very 
low energy and long lifetime, current estimates are around  $7.8$ eV 
and seconds to hours respectively. This nuclear transitions therefore 
is a promising candidate for a novel type of frequency standard and severly 
groups worldwide have set out to investigate this system. Our aim is 
to construct a "solid state nuclear clock", i.e. a frequency standard where 
Thorium ions are implanted into Calciumfluoride crystals transparent in vacuum
ultraviolet range. As a first step towards an accurate determination of the 
exact energy and lifetime of this isomer state we perform low-resolution
fluorescent spectroscopic measurements.
\end{abstract}        

\section*{Introduction}
Quantum frequency standards find a number of applications in fundamental science and applied technology. Nowadays the unit of time (SI second) is defined as 9,192,631,770 periods of the radiation of the ground state hyperfine transition
in $^{133}$Cesium~\cite{Markowitz58}. Cesium fountain primary frequency standards have reached an accuracy level of $3 \cdot 10^{-16}$~\cite{Parker10}.
At the same time, the progress in atom cooling and trapping together with sub-herz linewidth lasers allows to develop more accurate frequency standards based on optical transitions in neutral atoms and ions. For example, an accuracy level evaluation for a neutral $^{87}$Sr optical lattice clock is $1.5 \cdot 10^{-16}$~\cite{Ludlow08}, and the $^{27}$Al$^+$ quantum logic frequency standard demonstrated an inaccuracy level of order of $10^{-17}$~\cite{Chou10}.

An interesting candidate for the role of an etalon transition in quantum frequency standards is the isomer transition in a nucleus of $^{229}$Thorium.
This unique isotope has an extremely low-energy $^{229m}$Th isomeric state, expected at an optical wavelength of about 165\,nm.
The creation of a quantum frequency standards based on this transition will open new possibilities for experimental studies of basic fundamental laws of physics. The question ``are fundamental constants really constant" is among them. It is shown that the amplification factor between the drift of the nuclear transition frequency $\delta \omega/\omega$ and the drift of the fine structure constant $\delta \alpha/\alpha$ has an absolute value between $10^2$ and $10^4$~\cite{Litvinova09}. Hence the Thorium quantum frequency standard will be several order of magnitude more sensitive to the possible drift of $\alpha$ than existing standards based on the transitions in electronic shell of atoms or ions.

This paper is organized as follows. In section~\ref{sec:isomer} we review current knowledge on the isomer transition energy and lifetime. In section~\ref{sec:possibilities} we briefly discuss two main experimental approaches to the direct observation of the isomer transition: trapped ions and Thorium-doped crystals. In section ~\ref{sec:spectroscopy} we describe fluorescence spectroscopy of our Thorium-doped Calciumflouride crystals and estimate the optical pumping rate. Section~\ref{sec:future} gives an outlook on future steps.
\section{Isomer transition in the $^{229}$Th nucleus}
\label{sec:isomer}

\subsection{Isomer state energy}

The isomer transition energy in $^{229}$Th nucleus is currently known only from undirect measurements. In 1976, Kroger and Reich~\cite{Kroger76} studied the $\gamma$-ray spectrum resulting from the decay of $^{233}$Uranium. They concluded that the $^{229}$Th nucleus has a $J^{\pi}=3/2^+$ isomer state lying within $100$ eV above the $J^{\pi}=5/2^+$ ground state level.
The development of high quality germanium detectors (resolution from 300 to 900\,eV) allowed Helmer and Reich to measure more precise $\gamma$-energies in 1989 -- 1993 and to predict the energy of the nuclear transition to be $E=3.5 \pm 1.0$ eV~\cite{Reich90, Helmer94}. 

This unnaturally low value triggered a multitude of investigations, both theoretically and experimentally, trying to determine the transition energy precisely, and to specify other properties of the $J^{\pi}=3/2^+$ excited state of $^{229}$Th (such as lifetime and magnetic moment). 
However, direct searches for photon emission from the low-lying excited state have failed to observe a signal~\cite{Irwin97, Richardson98,
Shaw99, Utter99}. 
In 2005, Guim\~araes-Filho and Helene~\cite{Filho05} re-analysed the old experimental data and reported $E=5.5 \pm 1.0$ eV.

New indirect measurements with the most advanced X-ray microcalorimeter (resolution from 26 to 30\,eV) were performed by Beck et.al in 2007~\cite{Beck07}. 
They published a new value for the transition energy $E=7.6 \pm 0.5$ eV for the isomer nuclear transition, shifting it into the vacuum ultraviolet domain. 
This shift probably explains the failure in observing the transition in previous experiments. 
In 2009 Beck et. al. reanalyzed their results taking into account the non-zero probability of the $42.43 \; \mathrm{keV} \rightarrow \,
^{229m}\mathrm{Th}$ transition rate (estimated as 2\%) and published a revised version $E=7.8 \pm 0.5$ eV~\cite{Beck09}. 

The value $E=7.8 \pm 0.5$ eV is currently most accepted by the community but can not be considered definite untill
the direct measurements are performed successfully.
The dominant uncertainty in the prediction of Beck et.al is connected with the value of the branching ratio $b=1/13$~\cite{Beck07} from the 29.19 keV level to the ground state \cite{Sakharov10}. Estimation of  this value vary in different works. Sakharov~\cite{Sakharov10} mentions two alternative values for this branching ratio: 25\,\% in~\cite{Barci03} which would result in $E=9.3 \pm 0.6$ eV and 51\,\% in~\cite{Filho05} leading to $E=14.0 \pm 1.0$ eV.

\subsection{Isomer state lifetime}

While the direct measurement of the $3/2^+ \rightarrow 5/2^+$ ultraviolet transition in $^{229}$Th is still not performed, the isomer state lifetime $\tau$ remains unknow. Most recent estimations of the half life of the isomer state in a bare nucleus based on theoretical calculations of the matrix element of the magnetic moment for the transition between ground and isomer levels  were performed in~\cite{Ruchowska06}. 
The theory was checked by comparison with experimental data for transitions at higher energies. 
They found the half-life of this transition $T_{1/2}=(10.95 \mathrm{h})/ (0.025 E^3)$, where $E$ is given in eV. 
This corresponds $T_{1/2}=55$ min for $E=7.8$ eV.

For a Thorium atom or low charged ion the isomer lifetime can be significantly reduced. 
In work of Tkalya~\cite{Tkalya00} it is predicted that the probability of spontaneous magnetic dipole emission in a transparent non-magnetic dielectric medium with refractive index $n$ is $n^3$ times higher than in vacuum. 
The reason for this factor is the renormalisation of the density of emitted photon states. 
For a CaF$_2$ crystal, $n=1.55$ at wavelength $\lambda=159$ nm (corresponding $E=7.8$ eV), which gives a reduction factor of 3.7 for the lifetime. 

A more significant lifetime reduction can be connected with electronic bridge and bound internal conversion effects, i.e. isomer state decay can be amplified
by interaction between the nucleus and the electron shell. 
There are a number of theoretical calculations of electronic bridge processes. 
In an isolated neutral Thorium atom the isomer state lifetime is about $10^{-5}$\,s for $E=7.6$\,eV and about 4.5\,min for $E=3.5$\,eV~\cite{Karpeshin07}. 
In the Th$^{3+}$ ion for $E=7.6$\,eV the lifetime remains the same if the valence electron is in the ground state, and decreases 20 times if the valence electron
is in the metastable $7s$ state~\cite{Porsev10_3}. 
Similar calculations for the Th$^{4+}$ ion which is most likely to occur in the solid-state approach have yet to be performed.
In the Th$^{+}$ ion the lifetime decreases $10^2-10^3$ times for $E=3.5$ and $E=5.5$\,eV~\cite{Porsev10}, there is not sufficient data about the Th$^{+}$ ion spectrum
to calculate the lifetime at $E=7.8$\,eV.

Experimental attempts to measure the $^{229m}$Th lifetime were performed in~\cite{Inamura09, Kikunaga09}. 
These experiments are based on the theoretical prediction~\cite{Tkalya00PRC} that the $\alpha$ decay
rate for $^{229m}$Th is 2 to 4 times higher than for the ground state and that the energy spectrum of the $\alpha$
particles slightly differs for ground and isomer state. 

In~\cite{Inamura09} Thorium was electrodeposited onto a hollow cathode and then electronically excited by electric discharge. 
With some probability the excitation of the electronic shell was transfered to the nuclei~\cite{Karpeshin96} and therefore some amount of $^{229m}$Th was
obtained. 
After the end of the discharge the $\alpha$ activity of the sample was monitored and $\alpha$ counts 
with different energies were written to different channels to record a spectrum.
It was shown that $\alpha$ particles from Thorium are well energy separated from 
those of Thorium daughter products.
The excitation and detection procedure was repeated several times 
and a decaying component with half-life $T_{1/2}=2\pm 1$ min was observed in counts for 
$\alpha$ particles emitted from Thorium.
In the same work also plots indicating a time dependence of count ratios for $\alpha$ particles from daughter products are presented~\cite{Inamura09}. 
These plots demonstrate some time dependence higher than $3 \sigma$ variation (total number of counts is about 8000 per minute whereas the variation is about 300 per
minute). This could indicate some technical artifact being responsible for the observed decay of the $\alpha$ count rate from Thorium.

In the work of Kikunaga et.al~\cite{Kikunaga09} a sample of $^{233}$U was chemically purified from Thorium, then it was left for 1\,h to allow the growth of $^{229}$Th in its ground and isomer state, then the Thorium was extracted. 
The $\alpha$ spectrum from this fresh Thorium sample was recorded during the first 6000\,s and the following 6000\,s. 
They compared the $\alpha$ counts in the region of 4915--4955\,keV, where the $\alpha$ particles are produced in the decay of the $^{229m}$Th isomer state with much higher probability than in the decay of the ground state. 
Slightly more $\alpha$ counts where detected for the first time interval and the authors derived $T_{1/2}<2$\,h with $3\sigma$ confidence.
This experiment was not suited to measure half-lifes on the order of minutes because of the long chemical preparation process.

In general the analysis of these experiments is further complicated by the presence of various chemical compositions of Thorium with different decay rates depending on the chemical state (mainly ThO$^+$, Th and Th$^+$ in~\cite{Inamura09} and hydroxide and chloride complexes in~\cite{Kikunaga09}). So far, no clear evidence for $\alpha$ particles or photons originating from the  $^{229m}$Th isomer state has been presented. Therefore both the energy and the half-life of the $^{229m}$Th isomer state are still not determined definitely. 

\section{Approaches to direct measurements}
\label{sec:possibilities}
There are several groups worldwide working towards a direct measurements of the Thorium isomer transition.
(T. Schumm, Institute of Atomic and Subatomic Physics TU Vienna, Austria; E. R. Hudson, 
University of California, USA E. Peik, PTB, Germany; M. Chapman and A. Kuzmich, Georgia Institute of Technology, USA).
The expected excitation energy of 7.8\,eV enforces work with Thorium in an ionized state to avoid overwhelming optical response by the electron shell which would mask the nuclear signal.
Two main experimental directions are proposed. 
The first is the spectroscopy of Thorium ions in ion traps~\cite{Peik09, PorsevPeik10, Campbell09}. 
The second is the {\em solid-state nuclear spectroscopy}, where Thorium ions are implanted into a crystal (such as CaF$_2$ or
 LiCaF) transparent for vacuum ultraviolet light~\cite{Schumm, Rellergert10}. 
 
The main advantage of solid-state nuclear spectroscopy is the possibility of simultaneously excitating a macroscopic number (just a few orders of magnitude smaller than total number of atoms in a crystal) of Thorium nuclei whereas for ion trap spectroscopy this number is limited by $10^4 \div 10^5$. 
This high number facilitates the first search for the transitions compared to ion traps. 
The second  advantage is the fact that the experimental setup does not require any bulky equipment for ion trap creation such as ultrahigh vacuum and laser cooling. However, the solid-state spectroscopy method is not free of additional difficulties connected with the hardly removable or irremovable systematic effects in a crystal. 
Here, trapped ions present a very clean system with well-controllable parameters, probably ultimately more relevant for the realization of a high-performance laboratory
frequency standard. 

In our study we plan to first perform low-resolution fluorescence spectroscopy using CaF$_2$ as a host crystal to find the transition with 1 nm precision. 
CaF$_2$ has a large (more than 10\,eV) band gap. 
Th$^{4+}$ ions can replace Ca$^{2+}$ ions in the crystal lattice (ionic radii are 94\,pm for Th$^{4+}$ and 114\,pm for  Ca$^{2+}$), charge is compensated either by a Ca vacancy or two F interstitials. 

CaF$_2$ is not a good scintillator at the room temperature which is important as $^{229}$Th nuclei undergo $\alpha$ decay with a half-life of $7.8 \cdot 10^3$ years ($\alpha$ decay rate $\Gamma_{\alpha}=2.8 \cdot 10^{-12}$\,s). 
According to~\cite{Mikhalik06}, the total time of a photon scintillation burst following an $\alpha$ decay does not exceed 20\,$\mu$s. 
The scintillation decay constant for events caused by $\gamma$ or $X$ ray accompanying most of the $\alpha$ decays
is 2 times longer therefore the total decay time $t_{sc}$ should not exceed 50\,$\mu$s. $^{229}$Th has 8 short-living radioactive daughter elements. Therefore the number $n_{sc}$ of decays per second in the sample containing $N_{Th}$ Thorium nuclei with equilibrium number of daughter nuclei is about
\begin{equation}
n_{sc} = 8 N_{Th} \Gamma_{\alpha}\simeq 2.3 \cdot 10^{-11} N_{Th}.
\end{equation}
The scintillation will perturb the detection system during the part $n_{sc}\cdot t_{sc}$ of data aquisition time.
The requirement $n_{sc}\cdot t_{sc} \ll 1$ gives the upper limits for the total number of Thorium nuclei
in the sample: $N_{Th} \ll 8.7 \cdot 10^{14}$ for $t_{sc}=50 \, \mu$s.

Another important question concerns the position of possibly emerging electronic states connected with Thorium impurities in the crystal. 
If there is a level with an energy near the isomer transition energy, it can increase the rate of isomer state decay via the electronic bridge process. 
This critical point is issue of ongoing theoretical and experimental work.

\section{Fluorescent spectroscopy}
\label{sec:spectroscopy}
Low-resolution fluorescence spectroscopy is the first step towards an accurate determination of the exact energy and lifetime of the $^{229m}$Th isomer state.
\begin{figure}
\begin{center}
\resizebox{0.85\textwidth}{!}
{ \includegraphics{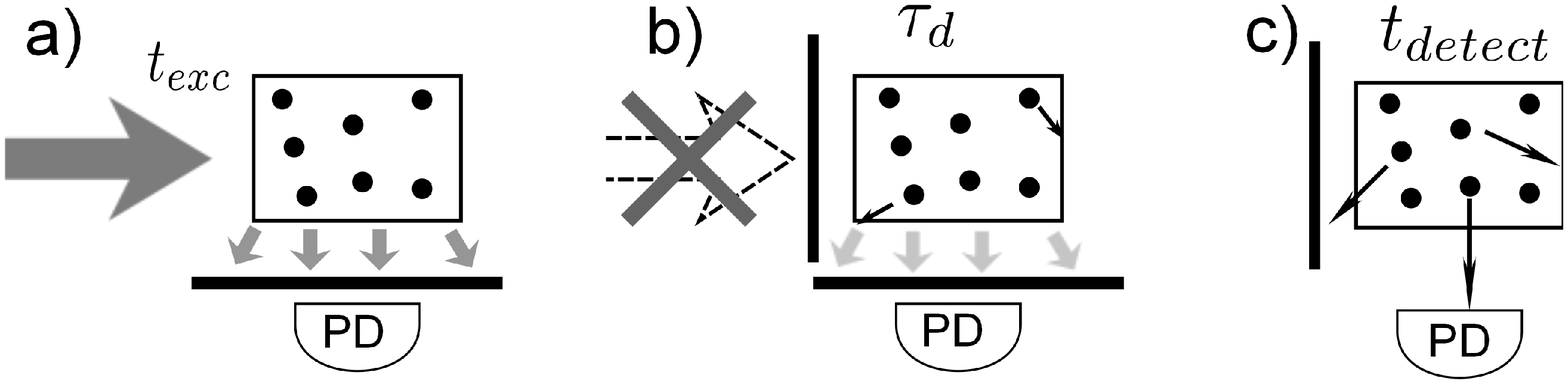}}
\end{center}
\def\baselinestretch{1.1}
\caption{Scheme of fluorescence spectroscopy procedure. It has three steps: a) irradiation of the sample
by vacuum ultraviolet light; b) decay of own crystal fluorescence; c)
count of the long-lived fluorescence}
\def\baselinestretch{1.5}
\label{fig:f1}
\end{figure}
Figure~\ref{fig:f1} shows a scheme of the planned spectroscopy experiment.
We plan to irradiate the sample for a time  $t_{exc}$ with the probe radiation (laser or synchrotron). 
We wait for a dead time $\tau_d$ to let the crystal fluorescence decay and then collect the fluorescence photons during time $t_{detect}$. 
Scanning the frequency of radiation and analysing the obtained photon counts, we plan to find the frequency of the isomer transition.

A recent study by the Hudson group using the synchrotron radiation available at the Advanced Light Source showed that typical relaxation time of crystal fluorescence (not connected to the isomer decay) for similar crystals (LiCAF, LiSAf NaYF etc.) doped with $^{232}$Th is less or about 1\,s~\cite{Hudson10}. 
We plan to perform similar measurements for $^{232}$Th-doped CaF$_2$ crystal using a high power laser system at the Max Born Institute in Berlin~\cite{MBI2007}.
As long as the isomer state lifetime considerably exceeds the crystal fluorescence lifetime and the isomer transition lies within the band gap of CaF$_2$ it should be possible to determine the isomer transition energy using the solid-state approach.

The number of photons detected  from a spontaneous nuclear isomer decay is:
\begin{equation}
N_{\gamma}=\frac{N_{Th}W}{\Gamma}\left(1-e^{-\Gamma t_{exc}}\right) 
\times e^{-\Gamma \tau_d} \times k \frac{\Omega \,\Gamma_s n^3}{4\pi \Gamma} 
\left(1-e^{-\Gamma t_{detect}}\right),
\end{equation}
where $\Gamma$ is the total decay rate, $W$ is the optical pumping rate, $\Gamma_s$ is the radiative 
decay rate of the bare nucleus, $n$ is the refraction index, $\Omega$ is the solid angle covered by the detector, $k$ is the detector efficiency. 

We now estimate the optical pumping rate $W$. For the low-resolution spectroscopy stage, the
spectrum of the radiation source will be broad and smooth, therefore $W$ depends only on the
spectral power density $dI/d\omega$ of the radiation at the transition frequency $\omega=\omega_{eg}$. 
Let us consider for the sake of simplicity the Lorentzian spectral density of intensity:
\begin{equation}
\frac{dI}{d\omega}=\frac{dI_0}{d\omega} \frac{1}{1+4(\omega-\omega_{eg})^2/\Gamma_L^2},
\end{equation}
where $\Gamma_L$ is the spectral linewidth of the source, $\frac{dI_0}{d\omega}$ is the spectral intensity density on resonance.

Th$^{4+}$ ion has a Rn-like electronic shell (noble gas) with zero angular momentum and zero spin. 
In the following we neglect the structure of the electronic shell as well as higher levels of the Th nucleus. 
Ground and isomer state are degenerate with respect to projections $m$ and $m'$ of angular momenta $J$ and $J'$ respectively (here we neglect
the interaction with ramdomly oriented spins of surrounding F nuclei). 
The magnetic field $\vec{B}$ of the spectroscopy radiation can be represented as $\vec{B}=\vec{B}_0 \, \cos(\omega t+\varphi(t))$, where
$\varphi(t)$ describes the random phase variations responsible for non-zero spectral linewidth 
$\Gamma_L$. Note that for the low-resolution spectroscopy $\Gamma_{L}$ just a few orders of magnitude smaller than $\omega$.

The Hamiltonian $\hat{H}$ in resonance approximation can be represented as
\begin{equation}
\hat{H}=\hat{H}_0 + \hbar \sum_{e,g} V_{eg}\left[|g\rangle\langle e|e^{i(\omega t + \varphi(t))}+
|e\rangle\langle g|e^{-i(\omega t + \varphi(t))} \right]+\hat{H}'.
\end{equation}
Here $V_{eg}=B_{0q} \mu_{eg}^q/(2\hbar)$, $B_{0q}$ is the $q$th covariant component of $\vec{B}_0$,
$\mu_{eq}^q$ is the $q$-th contravariant component of the matrix element $\vec{\mu}_{eq}$ of the nucleus 
magnetic momentum operator in cyclic coordinates, $q=m'-m$.
$\hat{H}_0$ is the bare Hamiltonian of the Thorium nucleus, $\hat{H}'$ describing the interaction of
the nucleus with all other fields except the spectroscopy radiation. 
We can choose a quantization axis, for example, along the radiation propagation direction, and classify excited states $e$
and ground states $g$ by angular momentum projection onto this axis. 
The set of equations for the density matrix $\rho_{ij}$ in the rotating frame with relaxation terms reads as:
\begin{align}
\dot{\rho}_{ee}&=-i\sum_g V_{eg} \left[\rho_{ge}-\rho_{eg}\right]-\Gamma \rho_{ee}+\sum_{e'}\Gamma_{||}^{ee'}\rho_{e'e'}; \displaybreak[1] \\
\dot{\rho}_{ee'}&=-i\sum_g \left[V_{eg} \rho_{ge'}-V_{e'g} \rho_{eg}\right]-(\Gamma+\Gamma_\perp^e) \rho_{ee'}, \quad e\neq e'; \displaybreak[1] \\
\dot{\rho}_{eg}&=(i\Omega-\Gamma')\rho_{eg}-i\sum_{g'}V_{eg} \rho_{g'g}+i\sum_{e'}V_{e'g}\rho_{ee'}-\Gamma'; \displaybreak[1] \\
\dot{\rho}_{gg}&=-i\sum_e V_{eg} \left[\rho_{eg}-\rho_{ge}\right]+\sum{e}\Gamma_{eg}\rho_{ee}+ \sum_{g'}\Gamma_{||}^{gg'}\rho_{g'g'};  \displaybreak[1] \\
\dot{\rho}_{gg'}&=-i\sum_e \left[V_{eg} \rho_{eg'}-V_{eg'} \rho_{ge}\right]-\Gamma_\perp^e \rho_{gg'}, \quad g\neq g'.
\end{align}
Here $\Gamma$ is the decay rate of the isomer state, $\Gamma_{\perp}^{e,g}$ and $\Gamma_{||}^{i,j}$ are decoherence
and mixing rates between magnetic states of the same energy level, $\Gamma'$ is the relaxation rate of 
optical coherence $\rho_{eg}$. 
We can estimate $\Gamma_{\perp}^{e,g}$ and $\Gamma_{||}^{i,j}$ via the spin-spin
relaxation time $\tau \simeq 20\,\mu$s for CaF$_2$ \cite{Jeener67}. Therefore, $\Gamma_{||}^{i,j} \approx
\Gamma_{\perp}^{e,g} \approx 1/\tau \approx 5\cdot 10^4$\,s$^{-1}$.
The optical coherence relaxation rate $\Gamma'$ can be represented as $\Gamma'=\Gamma_{L}/2+\Gamma/2+\Gamma_{ab}$ \cite{Mazets92}, where $\Gamma_{ab}$ is the relaxation rate due to adiabatic perturbation. 
For any reasonable low-resolution radiation source we can take
\begin{equation}
V_{eg} \ll \Gamma \ll \Gamma_{||}^{i,j} \approx \Gamma_{\perp}^{e,g} \ll \Gamma' = \Gamma_L/2.
\end{equation}
From (5-10) and the normalization condition $\sum_i \rho_{ii}=1$ we find:
\begin{equation}
\rho_{gg'}=\frac{\delta_{gg'}}{2J+1}; \quad \rho_{ee'} \simeq 0,\, e\neq e';
\end{equation}
\begin{equation}
\dot{\rho}_{exc}= W - \Gamma \rho_{exc};\quad W=\frac{1}{\Gamma'}\sum_{e,g}\frac{2 V_{e,g}^2}{2J+1}.
\end{equation}
where $\rho_{exc}=\sum_{e}\rho_{ee}$ is the total excited state population. 

Now we should calculate $\sum_{e,g}V_{eg}^2$. 
For the sake of simplicity we suppose that the radiation has a certain polarization along $q'$th cyclic unit vector, so $B_{0q}=|\vec{B}_{0}|\delta_{qq'}$.
In dipole approximation, $\Gamma = \frac{4 \omega^3}{3 \hbar c^3}\sum_g|\mu^q_{eg}|^2$. According the Wigner-Eckart theorem, 
$\mu_{eg}^q=||\mu||C_{Jm1q}^{J'm'}$, where $||\mu||$ is a reduced dipole moment, $C_{Jm1q}^{J'm'}$ is a
Clebsch-Gordan coefficient, $c$ is the light velocity in vacuum. 
Using $\sum_{m,q}\left|C_{Jm1q}^{J'm'}\right|^2=1$, we find
\begin{equation}
||\mu||=\sqrt{\frac{3\hbar c^3 \Gamma}{4 \omega^3}}.
\end{equation}
On the other hand,
\begin{equation}
\sum_{e,g} V_{eg}^2=\frac{\vec{B}_0^2 ||\mu||^2}{4 \hbar^2} \sum_{m,m'} \left|C_{Jm1q'}^{J'm'}\right|^2
=\frac{\vec{B}_0^2 ||\mu||^2}{4 \hbar^2} \frac{(2J'+1)}{3}.
\end{equation}

The amplitude $|\vec{B}_0|$ of magnetic field is easy to express via the total radiation intensity $I$: 
$|\vec{B}_0|=8 \pi I/c$. Integrating (3) over $\omega$, we obtain $I=\pi \Gamma' \frac{dI_0}{d\omega}$,
therefore
\begin{equation}
|\vec{B}_0^2|=\frac{8\pi^2\Gamma'}{c}\frac{dI_0}{d\omega}.
\end{equation}
From (12)-(15) we finally obtain the expression for the optical pumping rate $W$:
\begin{equation}
W=\frac{2}{3} \; \frac{c^2 \Gamma \pi^2}{\hbar \omega^3}\; \frac{dI_0}{d\omega}
=\frac{\lambda^5\Gamma}{24 \pi^2 c^2 \hbar}\frac{dI_0}{d\lambda}.
\end{equation}
Here we took into account the values of total angular momentum $J=5/2$ and $J'=3/2$ in the ground and
excited state respectively. 
Typical values of $W$ are estimated in Table 1. 
We see that if the sample contains $N_{Th}=10^{13}$ Thorium nuclei, and if we use a powerful excitation source (e.g. ALS synchrotron), we can
excite up to $3.25 \cdot 10^4$ nuclei per second.

\newcommand{\ST}{\vphantom{$\displaystyle{\frac{a}{b}}$}}
\begin{table}[t]
\caption{Estimation of optical pumping rate $W$ for differnt sources}
\label{tab:freqlife}
\begin{center}
\begin{tabular}{|l|c|c|c|}\hline
\ST Source & $\lambda$, nm &$dI/d\lambda$, W$/$cm$^2\cdot$nm & $W$, s$^{-1}$ \\ \hline
\ST McPherson 632 thermal lamp& 113-200 & $2.5\cdot 10^{-6}$ (165 nm) & $6 \cdot 10^{-16}$ \\ \hline
\ST Q-switched laser (NT342B-SH) &210--420 & 4 (225 nm) & $10^{-9}$\\ \hline
\ST Synchrotron (ALS, USA) &40--250& 32 (165 nm) & $3.25\cdot 10^{-9}$\\ \hline
\end{tabular}
\end{center}
\end{table}
\def\baselinestretch{1.5}
%

\section{Future plans}
\label{sec:future}
We are currently in preparation for the low-resolution spectroscopy stage. 
Our immediate plan is to develop and improve noise reduction techniques and to measure the fluorescence rate of a $^{232}$Th-doped
CaF$_2$ crystal after irradiation with a high-power laser field to accurately determine the crystal fluorescence time $\tau_d$. 
In parallel we perform absorption measurments on $^{232}$Th-doped to determine whether addtional electronic states within the band gap are created by the doping process. Theoretical simulations of the microscopic properties of the doping complex are under way.
These measurements will help us to determine the optimal measurement strategy concerning the timings $t_{exc}$, $\tau_d$, and $t_{detect}$ as well as the required excitation source and the optimal doping concentration.

In a second step we plan to grown the $^{229}$Th-doped CaF$_2$ crystal and perform the low-resolution spectroscopy measurements as discussed above.

Once the isomer energy will be found, we will proceed to the next step, the high-resolution spectroscopy.
We plan to use high-harmonic generation (5th harmonic of 800\,nm) of a femtosecond frequency comb system to generate a narrow-linewidth excitation source with absolute frequency callibration. At this stage, we will be able to quantify line shifts and broadenings due to the crystal environment and validate the general experimental approach.

Finally we plan to actively stabilize the frequency comb to the nuclear transition and thereby realize to a solid-state nuclear frequency standard.

\section*{Acknowledgements}
The research was funded by the European Research Council (ERC): starting grant 258603 – NAC,
and by the Austrian Science Fund (FWF): M1272-N16 (TheoNAC) and Y481-N16.

We thank E. Peik, Chr. Tamm, and E. Hudson for insightful discussion and 
support. We thank R. Uecker, R. Rabe, and R. Bertram from IKZ Berlin for 
growing the $^{232}$Th-doped CaF$_2$ crystals and performing the ICP-MS; 
furthermore we acknowledge the support by J. Friedrich and P. Berwian 
from IISB Erlangen. R. Jackson we thank for crystal impurity 
simulations. We thank C. Wickleder and M. Adlung for optical 
characterization and advice concerning color centers.



\begin{thebibliography}{99}\setlength{\itemsep}{0mm}\setlength{\itemsep}{0mm}
{
\bibitem{Markowitz58} W.~Markowitz, R.~Glenn Hall, L.~Essen and 
J.~V.~L.~Parry, Phys. Rev. Lett. {\bf 1}, 3, 105-107 (1958).

\bibitem{Parker10}T.~E.~Parker, Metrologia {\bf 47}, 1-10 (2010).

\bibitem{Ludlow08}A.~D.~Ludlow, T.~Zelevinsky, G.~K.~Campbell, et.al.,
Science {\bf 319}, 1805-1808 (2008).

\bibitem{Chou10}C.~W.~Chou, D.~B.~Hume, J.~C.~J.~Koelemeij,
et.al., Phys. Rev. Lett. {\bf 104}, 070802 (2010).

\bibitem{Litvinova09}E.~Litvinova, H.~Feldmeier, J.~Dobaczewski, V.~Flambaum, 
Phys. Rev. C {\bf 79}, 064303 (2009); arXiv:0901.1240.

\bibitem{Kroger76}L.~A.~Kroger and C.~W.~Reich, Nuclear Physics A {\bf 259}, 29-60 (1976)

\bibitem{Reich90}C.~W.~Reich and R.~G.~Helmer,
Phys. Rev. Lett. {\bf 64}, 271-273 (1990)

\bibitem{Helmer94}R.~G.~Helmer and C.~W.~Reich,
Phys. Rev. C{\bf 64}, 1845-1858 (1994)

\bibitem{Irwin97}G.~M.~Irwin, K.~H.~Kim, Phys. Rev. Lett. {\bf 79}, 990-993 (1997)

\bibitem{Richardson98}D.~S.~Richardson, D.~M.~Benton, D.~E.~Evans, et.al.,
Phys. Rev. Lett. {\bf 80}, 3206-3208 (1998)

\bibitem{Shaw99}R.~W.~Shaw, J.~P.~Young, S.~P.~Cooper, and O.~F.~Webb.
Phys. Rev. Lett. {\bf 82}, 1109-1111 (1999)

\bibitem{Utter99}S.~B.~Utter, P.~Beiersdorfer, A.~Barnes, et.al.,
Phys. Rev. Lett. {\bf 82}, 505-508 (1999)

\bibitem{Filho05}Z.~O.~Guim\~araes-Filho, O.~Helene,
Phys. Rev. C 71, 044303, 4 p. (2005)

\bibitem{Beck07}B.~R.~Beck, J.~A.~Becker, P.~Beiersdorfer, G.~V.~Brown, et.al.,
Phys. Rev. Lett. {\bf 98}, 142501 (2007).

\bibitem{Beck09}B.~R.~Beck, C.~Y.~Wu, P.~Beiersdorfer, et.al.,
12th International Conference on Nuclear Reaction Mechanisms, Varenna, Italy,
LLNL-PROC-415170 (2009).

\bibitem{Sakharov10} S.~L.~Sakharov. {\em On the Energy of the 3.5-eV Level in $^{229}$Th}.  
Physics of Atomic Nuclei {\bf 73}, No. 1,  pp. 1–8 (2010).

\bibitem{Barci03} V.~Barci, G.~Ardisson, G.~Barci-Funel, et.al., Phys. Rev.C {\bf 68}, 034329 (23pp) (2003)

\bibitem{Ruchowska06} E.~Ruchowska, W.~A.~P\l\'ociennik, J.~\.Zylicz, et.al., Phys. Rev. C {\bf 73}, 044326 (2006)

\bibitem{Tkalya00} E.~Tkalya, JETP Lett. {\bf 71}, 311-313 (2000)

\bibitem{Karpeshin07} F.~F.~Karpeshin and M.~B.~Trzhaskovskaya, Phys. Rev. C {\bf 76}, 054313 (2007)

\bibitem{Porsev10_3} S.~G.~Porsev and V.~V.~Flambaum, Phys. Rev. A {\bf 81}, 032504 (2010)

\bibitem{Porsev10} S.~G.~Porsev and V.~V.~Flambaum, Phys. Rev. A {\bf 81}, 042516 (2010)

\bibitem{Inamura09}T.~T.~Inamura and H.~Haba, Phys. Rev. C {\bf 79}, 034313 (2009) - 10 p.

\bibitem{Kikunaga09}H.~Kikunaga, Y.~Kasamatsu, H.~Haba, et.al., 
Phys. Rev. C {\bf 80}, 034315 (2009) - 4 p.

\bibitem{Karpeshin96} F.~F.~Karpeshin, I.~M.~Band, M.~B.~Trzaskovskaya, M.~A.~Listengarten,
Phys. Lett. B {\bf 372}, 1-7 (1996).

\bibitem{Tkalya00PRC}E.~V.~Tkalya, A.~N.~Zherikhin, V.~I.~Zhudov,
Phys. Rev. C {\bf 61}, 064308 (2000) - 5 p.

\bibitem{Peik09}E.~Peik, K.~Zimmermann, M.~Okhapkin, Chr.~Tamm
Proceedings of the 7th Symposium on Frequency Standards and Metrology (2009), arXiv:0812.3458

\bibitem{PorsevPeik10} S.~G.~Porsev, V.~V.~Flambaum, E.~Peik, and Chr.~Tamm,
Phys. Rev. Lett. {\bf 105}, 182501 (2010)

\bibitem{Campbell09} C.~J.~Campbell, A.~V.~Steele, L.~R.~Churchill, M.~V.~DePalatis, et.al.,
 Phys. Rev. Lett. {\bf 102}, 233004 (4pp) (2009)

\bibitem{Schumm} T. Schumm, {\em see www.thorium.at}

\bibitem{Rellergert10}W.~G.~Rellergert, D.~DeMille, R.~R.~Greco, et.al., 
Phys. Rev. Lett. {\bf 104}, 200802 (4pp) (2010)

\bibitem{Mikhalik06}V.~B.~Mikhailik, H.~Kraus, J.~Imber, D.~Wahl, 
Nucl. Instr. Meth. Phys. Res. A {\bf 566}, 522-525 (2006)

\bibitem{Hudson10}W.~G.~Rellergert, S.~T.~Sullivan, D.~DeMille, et.al., arXiv:1011.0769, 
Proc. 11th Europhysical Conference on Defects in Insulating Materials (2010); 

\bibitem{MBI2007}I.~P. Tzankov, O. Steinkellner, J. Zheng, et.al., Optics Express {\bf 10}, 6389 (2007)

\bibitem{Jeener67}J.~Jeener, P.~Broekaert, Phys. Rev. {\bf 157}, 232 (1967)

\bibitem{Mazets92}I.~Mazets, B.~Matisov, Sov. phys. JETP {\bf 74}, 13 (1992)
}


\end{thebibliography}
\end{document}